\documentclass[11pt,a4paper]{article}
\usepackage[english]{babel}
\usepackage{amsmath,amsthm,amssymb,epsfig,latexsym}
\usepackage{color}
\usepackage{ulem}
\usepackage[numbers,square,sort&compress]{natbib}

\newcommand{\so}{\scriptscriptstyle \rm I}
\newcommand{\st}{\scriptscriptstyle \rm I\hspace{-1pt}I}

\newcommand{\bu}{\bar u}
\newcommand{\bv}{\bar v}
\newcommand{\bx}{\bar x}

\newcommand{\buc}{\bar{u}^{\scriptscriptstyle C}}
\newcommand{\bub}{\bar{u}^{\scriptscriptstyle B}}
\newcommand{\bvc}{\bar{v}^{\scriptscriptstyle C}}
\newcommand{\bvb}{\bar{v}^{\scriptscriptstyle B}}

\newcommand{\bet}{\bar\eta}

\newcommand{\be}[1]{\begin{equation}\label{#1}}
\newcommand{\ba}[1]{\begin{multline}\label{#1}}
\newcommand{\ee}{\end{equation}}
\newcommand{\ea}{\end{multline}}

\newcommand{\tr}{\mathop{\rm tr}}

\def\qed{\hfill\nobreak\hbox{$\square$}\par\medbreak}
 \makeatletter
 \@addtoreset{equation}{section}
 \makeatother
 
\newcommand{\bea}{\begin{eqnarray}}
\newcommand{\eea}{\end{eqnarray}}

\def\BB{{\mathbb{B}}}
\def\CC{{\mathbb{C}}}

\def\hBB{{\hat{\mathbb{B}}}}
\def\hCC{{\hat{\mathbb{C}}}}
\def\mT{{\mathcal{T}}}

\def\rvac{|0\rangle}
\def\lvac{\langle 0 |}

\newtheorem{theorem}{Theorem}[section]

\begin{document}

\vspace{12pt}

\begin{center}
\begin{LARGE}
{\bf New approach to scalar products of Bethe vectors}
\end{LARGE}

\vspace{40pt}

\begin{large}
{A.~Liashyk \footnote{E-mail: a.liashyk@gmail.com}}
\end{large}

 \vspace{12mm}

{\it Bogoliubov Institute for Theoretical Physics, NAS of Ukraine,  Kiev, Ukraine}

\vspace{4mm}

{\it National Research University Higher School of Economics, Faculty of Mathematics, Moscow, Russia}

\vspace{4mm}

{\it Skolkovo Institute of Science and Technology, Moscow, Russia}

\end{center}

\vspace{4mm}

\begin{abstract} 
 We consider quantum integrable models solvable by the algebraic Bethe ansatz and possessing $\mathfrak{gl}(2)$-invariant $R$-matrix. We study the models of both periodic boundary conditions and boundary conditions based on reflection algebra. We present a new method to calculate the scalar products based on formula of an action of transfer matrix of a model onto Bethe vector. Using this method we identify some coefficient in the multiple action of the transfer matrix with the scalar product between on-shell and off-shell Bethe vectors. This allows us to find determinant representation of the scalar products in both types of boundary conditions.
\end{abstract}

\section{Introduction} 

 One of the most fruitful method of research of quantum integrable systems is algebraic Bethe ansatz. 
This method is developed  by the Leningrad school (primarily L. Faddeev, L.  Tahtadjian, and E. Sklyanin) \cite{FadST79,FadT79,FadLH96}.
It is used for finding spectra and eigenvector of quantum Hamiltonians\cite{FadST79,FadT79,FadLH96}. 
Also it can be used for an efficient calculation of the form factors and correlation functions \cite{BogIK93L,KitMT00,KitKMST12,GohKS04}.

 Scalar products of Bethe vectors play a very important role in the algebraic Bethe ansatz. 
They are building blocks on the way of  calculating form factors and correlation functions within this framework. 
The first results about scalar products ware obtained by A. Izergin  and  V. Korepin \cite{IzeK84,Ize87}.
Then N. Slavnov found the determinant representation of the scalar products \cite{Sla89}. 
Later this result was formulated as the scalar product between on-shell off-shell Bethe vectors is proportional to determinant of matrix constructed by derivatives of the eigenvalues of generation function of Hamiltonians \cite{KitMT99}. 
M. Gaudin formulated a hypothesis about the norm of the Hamiltonian eigenfunction \cite{Gau72} and V. Korepin proved it for a wide class of quantum integrable models \cite{Kor82}.
According to this hypothesis, the square of the eigenfunction norm is proportional to a Jacobian closely related to the Bethe equations.  
  
  Usually algebraic Bethe ansatz connects quantum integrable systems with periodic boundary conditions. To describe systems with open boundary conditions Sklyanin  introduced reflection algebra  \cite{Skl88}. In the paper \cite{Kit07} one can find an expression of scalar products for spin chains with open boundary conditions.

 It was observed in \cite{KMST02} that the coefficients in the multiple action of transfer matrix onto Bethe vector have determinant form. 
The main result of this paper is connection of one coefficient in the action with Slavnov formula of scalar product of Bethe vectors. 
It allows us a new way of calculation of the Slavnov formula, which is very useful.
Beside usual periodic case we repeat all result for open boundary case.

 This paper is organized as follows.
We recall description of $RTT$-algebra and reflection algebra in section~\ref{Notations}. 
There we also give a shorthand notation for sets of parameters and products over those parameters. 
In section~\ref{periodic} we briefly describe of Bethe vectors within ABA approach. Then we formulate and prove the main theorem of the paper in periodic case. 
Separate subsection contains derivation of the determinant representation of scalar product of Bethe vector (also known as the Slavnov formula). 
Section~\ref{reflect} repeats the logic of the previous one for reflection boundary conditions case.

\section{Defenition and notation}\label{Notations} 

\subsection{RLL-algebra}

 Let $R(u,v)$ be matrix associated with the vector representation of the Yangian $Y\left(\mathfrak{gl}_2\right)$:
\begin{equation}\label{rmatrix}
 R(u,v) = \mathbf{1}\otimes\mathbf{1} + g(u,v) \mathbf{P}, \quad  
 g(u,v) = \frac{c}{u-v},
\end{equation}
where $\mathbf{1}$ is the identity matrix in $\mathbf{C}^2$, $\mathbf{P}$ is the permutation matrix in $\mathbf{C}^2\otimes\mathbf{C}^2$, $c$ is a constant, and and $u, v$ are arbitrary complex parameters called spectral parameters.

 The key object of algebraic Bethe ansatz is monodromy matrix $T(u)$. 
Matrix elements of $T(u)$ are operator valued and act in a Hilbert space $\mathcal{H}$ which is a physical space of a quantum model. 
The main property of $T(u)$ is $RTT$-relation:
\begin{equation}\label{rtt}
 R(u,v) \;  T^{(1)}(u) \, T^{(2)}(v) = 
 T^{(2)}(v) \,  T^{(1)}(u) \; R(u,v), 
\end{equation}	
where $T^{(i)}(u)$ acts in $i$-th tensor multiplier. 
The simple consequence of the equation \eqref{rtt} is that the monodromy matrix entries commute with themselves for different arguments $[T_{ij}(u),T_{ij}(v)]=0$.

 The transfer matrix is defined as the trace of the monodromy matrix
\begin{equation}\label{tm}
 t(u) = \tr T(u).
\end{equation} 
One can easy check that $ [ t(u), t(v) ] = 0$ \cite{FadST79,FadT79,FadLH96} . Thus it allows us identify transfer matrix with generating function of integral of motion of an integrable system. Such models usually have periodic boundary conditions.

 We assume that the space $\mathcal{H}$ contain vacuum vector $\rvac$ such that
\begin{equation}\label{vac}
 \begin{aligned}
  T_{ii}(u) \rvac &= \lambda_i(u) \rvac, \\
  T_{21}(u) \rvac &= 0.
 \end{aligned}
\end{equation} 
Here $\lambda_i(u)$ are some complex-valued functions depending on concrete quantum integrable model. Multiple actions of $T_{12}$ with different arguments onto vacuum $\rvac$ generate a basis of quantum physical space $\mathcal{H}$ \cite{FadST79,FadT79,FadLH96} .
 
 \underline{Generalized model.} In the framework of this paper we assume that $\lambda_i(u)$ are free functional parameters and we do not specify any concrete their dependencies \cite{Kor82}. 
More or less it means that one can find concrete quantum integrable model for any specific choice of  $\lambda_i(u)$. 
For our purposes it is enough to assume that $\lambda_i$ and their derivatives do not depend on their arguments in direct way only through the Bethe equations \eqref{be} for on-shell Bethe vectors.

\subsection{Reflection algebra}

 To describe quantum spin chain with open boundary conditions Sklyanin introduced reflection algebra  \cite{Skl88}.
It is given by two following equations
\begin{multline}
  R_{12}(u_1-u_2) K_-^{(1)}(u_1)R_{12}(u_1+u_2)K_-^{(2)}(u_2) =  \\
  K_-^{(2)}(u_2)R_{12}(u_1+u_2) K_-^{(1)}(u_1)R_{12}(u_1-u_2)
\end{multline}
and
\begin{multline}
  R_{12}(-u_1+u_2) K_+^{(1)}(u_1)R_{12}(-u_1-u_2-2 c)K_+^{(2)}(u_2) =  \\
  K_+^{(2)}(u_2)R_{12}(-u_1-u_2-2 c) K_+^{(1)}(u_1)R_{12}(-u_1+u_2).
\end{multline}
In the framework of this paper we consider only diagonal solutions of reflection equation  on $K_{\pm}$.
General solution is 
\begin{equation}
 K_-(u) = K(u, \xi_-), \quad  K_+(u) = K(u+c, \xi_+),
\end{equation}
where
\begin{equation}
  K(u,\xi) = 
	\begin{pmatrix} 
      \xi + u  & 0 \\
      0 & \xi - u 
	\end{pmatrix}.
\end{equation}

 Now we can introduce reflection monodromy matrix \cite{Skl88}
\begin{equation}
  \mT (u) = T(u) \, K\left(u - \frac{c}{2} ,\xi_-\right)\, \sigma_2 \, T^t(-u) \, \sigma_2 
\end{equation}
and reflection transfer matrix 
\begin{equation}\label{rtm}
 \hat{t}\, (u) = {\rm Tr}\;  K\left(u + \frac{c}{2}, \xi_+ \right)\,   \mT (u).
\end{equation}

 One can prove \cite{Skl88} that
\begin{equation}
 [ \hat{t}\, (u) ,  \hat{t}\, (v) ] = 0
\end{equation}
and $ \hat{t}\, (u) $ can be identify with generating function of Hamiltonians.

 We use conception of generalized model as for periodic case too. 

 In this case we have the same vacuum as for periodic case \eqref{vac} with properties
\begin{equation}
 \begin{aligned}
  \mathcal{T} _{11}(u) \rvac &=   \left( u + \xi_-  - \frac{c}{2} \right) \lambda_1(u)\lambda_2(-u)  \rvac, \\
  \mathcal{T} _{22}(u) \rvac &= \left(\frac{c-2u}{2u}\left( u- \xi_-  + \frac{c}{2} \right) \lambda_1(-u)\lambda_2(u) + \frac{ u+ \xi_- - \frac{c}{2}}{2u}  \lambda_1(u)\lambda_2(-u) \right)  \rvac, \\
  \mathcal{T} _{21}(u) \rvac &= 0.
 \end{aligned}
\end{equation} 

Here $\lambda_i(u)$ are the same as in \eqref{vac}. Multiple actions of $\mT_{12}$ with different arguments onto vacuum $\rvac$ generate a basis of quantum physical space $\mathcal{H}$ for models with open boundary case \cite{Skl88}.

\subsection{Notation}

 In this paper we use our standard notation. 
Besides the function $g(u,v)$ in \eqref{rmatrix} we use different rational functions 
\begin{equation}
 \begin{aligned}
  f(u,v) &= \frac{u-v+c}{u-v}, \\
  h(u,v) &= \frac{u-v+c}{c}.
 \end{aligned}
\end{equation}

 We use a bar to denote sets of variables, for example, $\bu$, $\bv$. 
More precisely $\bu = \{u_1, \ldots, u_n\}$, where $n$ is a number of elements in  $\bu$.
Individual element of the set is denoted by Latin subscript, for instance, $u_i$ is an element of $\bu$, $v_j$ is an element of $\bv$ etc.
A notation $\bu_j$ is used for subsets complementary to the element $u_j$, that is, $\bu_j = \bu \backslash u_j$, $\bv_i = \bv \backslash v_i$, and so on.
We use notation $-\bu$ for set of variables with opposite sign $-\bu = \{-u_1,\ldots, - u_n\}$ and notation $\bu^2$ for set of squares of variables  $\bu^2 = \{u_1^2,\ldots, u_n^2\}$.
 
 Subsets of variables are denoted by roman indices: $\bu_{\so}$, $\bv_{\st}$, and so on. 
In particular, we consider partitions of sets into subsets. 
Then the notation $\{ \bu_{\so} , \bu_{\st}  \} \vdash \bu $ means that the set $\bu$ is divided into two disjoint subsets $\bu_{\so}$ and $\bu_{\st}$. 

 We use a shorthand notation for products of the rational functions $g$, $f$, and $h$.
Namely, if the rational functions have one or two sets of variables as argument, this means that one should take the product over all elements of the corresponding set (or the double product over elements of both sets). For example,
\begin{equation}
 f(z,\bu) = \prod_{u_i \in \bu} f(z, u_i), \quad
 f(\bu_{\so},\bu_{\st}) = \prod_{u_i \in \bu_{\so}}  \prod_{u_j \in \bu_{\st}} f(u_i, u_j).
\end{equation}
 
 We use the same prescription for the products of commuting operators and their eigenvalues
\begin{equation}
 T_{ij} (\bu) = \prod_{u_k \in \bu}  T_{ij} (u_k), \quad
 t(\bar z) = \prod_{z_i \in \bar z}  t(z_i) , \quad
 \lambda_{i} (\bv) = \prod_{v_j \in \bv} \lambda_{i} (v_j).
\end{equation} 

 Finally, by definition, any product over the empty set is equal to $1$.
A double product is equal to $1$ if at least one of the sets is empty. 

We also use notation $\Delta ' (\bu)$ and  $\Delta (\bu)$ 
\begin{equation}
 \Delta ' (\bu) =  \prod_{j<k} g(u_j,u_k), \quad
 \Delta (\bu) =  \prod_{j>k} g(u_j,u_k).
\end{equation}

\section{Periodic case}\label{periodic}

\subsection{Bethe vector}

 Now we pass to the description of Bethe vectors.
They belong to a Hilbert space  $\mathcal{H}$, in which the operators $T_{ij}(u)$ act. 
We do not specify this space, however, according to the \cite{FadST79,FadT79,FadLH96} in the framework of generalized model off-shell Bethe vector for set $\bu = \{u_1, \ldots, u_n\}$ is  
\begin{equation}\label{bv}
 \BB(\bu) = T_{12}(\bu) \rvac = T_{12}(u_1)\ldots T_{12}(u_n) \rvac.
\end{equation}
Due to the commutation relations \eqref{rtt} the Bethe vector \eqref{bv} is symmetric over permutations of the parameters $\bu$.

 One can prove that off-shell Bethe vector becomes eigenvector of transfermatrix $t(u)$ \eqref{tm} 
\begin{equation}
 t(z) \, \BB(\bu) = \tau(z | \bu) \, \BB(\bu)
\end{equation}
if parameters $\bu$ satisfy the system of Bethe equations \cite{FadST79,FadT79,FadLH96}
\begin{equation}\label{be}
 \frac{\lambda_1(u_i)}{\lambda_2(u_i)} = \frac{f(u_i,\bu_i)}{f(\bu_i,u_i)},
\end{equation}
with eigenvalue 
\begin{equation}\label{tme}
 \tau(z | \bu) = \lambda_1 (z) f(\bu,z) + \lambda_2(z) f(z,\bu).
\end{equation} 
 
 Dual Bethe vectors belong to the dual space $\mathcal{H}^*$.
They can be obtained as a transposition of the Bethe vectors \cite{FadST79,FadT79,FadLH96}.
We denote dual Bethe vectors by $\CC(\bu)$.
\begin{equation}
  \CC(\bu) =  \lvac T_{21}(\bu), 
\end{equation}
where $ \lvac$ is dual vacuum from the dual space $\mathcal{H}^*$ with properties
\begin{equation}
 \begin{aligned}
  \lvac T_{ii}(u)  &= \lambda_i(u) \lvac, \\
  \lvac T_{12}(u)  &= 0.
 \end{aligned}
\end{equation} 
Here $\lambda_i(u)$ are the same as in \eqref{vac}.
Dual Bethe vector becomes on-shell too, if the set $\bu$ satisfy the system of Bethe equations \eqref{be} with the same eigenvalue \eqref{tme}.

\subsection{Action of the transfer matrix and main theorem}

 In this subsection we present the main results of the paper for periodic boundary case. 
Here we give the theorem \ref{KAth} and its proof. 
This theorem  allows us to calculate scalar product of on-shell and off-shell Bethe vectors, knowing the action of transfer matrix onto a Bethe vector.

 Let us recall \cite{FadST79,FadT79,FadLH96} the formula of action of transfer matrix \eqref{tm} onto Bethe vector \eqref{bv}
\begin{multline} 
   t(z) \; \BB(\bu) =  \tau(z|\bu) \,\BB(\bu) + \sum_j g(z, u_j)\left( \lambda_1(u_j) f(\bu_j,u_j)  - \lambda_2(u_j) f(u_j,\bu_j)\right) \BB(\{z, \bu_j \}) = \\
  \tau(z|\bu) \,\BB(\bu) -  \frac{1}{c}  
  \sum_j g(z, u_j) \mathop{{\rm res}}\limits_{w=u_j} \tau(w|\bu)
 \; \BB(\{z, \bu_j \}).
\end{multline}
We see that the action is given only by eigenvalue of transfer matrix  $\tau$ \eqref{tme}. 

 Taking into account this action we can formulate the main theorem of our paper.
 
\begin{theorem}\label{KAth} 
Let us consider multiple action of a transfer matrix onto off-shell Bethe vector with  $\#\bu = \#\bv$
 \begin{equation}\label{act}
   t(\bv) \; \BB(\bu) = \sum R(\bv\,|\,\bet_{\so},\bet_{\st}) \; \BB(\bet_{\st}),
 \end{equation}
where $\bar\eta=\{\bu,\bv\}$. 
Here the sum is taken over partitions $\bet$ into two subsets $\bet_{\so}$ and $\bet_{\st}$, such that  $\#\bet_{\so} = \#\bet_{\st}$.

Then the coefficient of $\BB(\bv)$ of multiple action of the transfer matrixes  with $\#\bv=\#\bu$ coincides with Slavnov formula \cite{Sla89} for the scalar product of on-shell--off-shell Bethe vectors
 \begin{equation}\label{kat1}
  S(\bv | \bu) \equiv \CC(\bv) \BB(\bu) = R(\bv\,|\,\bu ,\bv).
 \end{equation}
 where set $\bv$ satisfies Bethe equations and set $\bu$ is free (off-shell).
\end{theorem}
{\sl Proof.}
Let a set $\bv$ satisfy Bethe equations, a set $\bu$ be free (off-shell), and a set $\bx$ be a set of free parameters. And let $\#\bv=\#\bu=\#\bx$.
Then one can consider an expression
\begin{equation}\label{ctb1}
  \CC(\bv)\; T(\bx)\; \BB(\bu).
\end{equation}
The left vector is the eigenvector of the transfer matrix
\begin{equation}\label{eig}
  \CC(\bv)\; T(x) = \tau(x|\bv) \;\CC(\bv).
\end{equation}
One can substitute equations \eqref{act} and \eqref{eig} in equation \eqref{ctb1}
\begin{equation}\label{ctb2}
   \tau(\bx|\bv)\, \CC(\bv)\BB(\bu) =  \sum R(\bx\,|\,\bet_{\so},\bet_{\st}) \; \CC(\bv) \BB(\bet_{\st}),
\end{equation}
where $\bet =\{\bx,\bu\}$.
And now one can put $\bx \to \bv$ in equation \eqref{ctb2}.
\begin{multline}
  \lim_{x\to v_i} \tau(x \,|\,\bv) =  - c\, \lambda_1'(v_i) f(\bv_i,v_i) + c\, \lambda_2'(v_i) f(v_i,\bv_i) + \\ \lambda_1(v_i) f(\bv_i,v_i)\left(1-\sum_{j\ne i} t(v_j,v_i)\right)+\lambda_2(v_i) f(v_i,\bv_i)\left(1+\sum_{j\ne i} t(v_i,v_j)\right) = \\
  - c\,   \lim_{x\to v_i} \tau(x \,|\,\bv) =  - c\, \lambda_1'(v_i) f(\bv_i,v_i) + c\, \lambda_1(v_i) f(\bv_i,v_i) X(v_i) + UT,
\end{multline}
where $X(v_i) = \left. \ln\left(\frac{\lambda_1(z)}{\lambda_2(z)}\right)'\right|_{z=v_i}$ and $UT$ does not contain derivatives.

It is clear that $S(\bv|\{\bv_{\so},\bu_{\st}\})$ can contain only $X(v_i)$ with $v_i \in \bv_{\so}$.

Let us compare coefficients of $X(\bv)$ in equation \eqref{ctb2}. In RHS one can get $X(\bv)$ from the norm of Bethe vector $S(\bv|\bv)$. An explicit expression one can find in the \cite{Gau72,Kor82}, but here we use only coefficient of $X(\bv)$
\begin{multline}
  (-c)^{\# \bv} \lambda_1(\bv) \prod_{i\ne j} f(v_i,v_j) X(\bv)\;  S(\bv|\bu)+\dots =
  R(\bv\,|\,\bu ,\bv)\, S(\bv | \bv) + \dots = \\
  R(\bv\,|\,\bu ,\bv) (-c)^{\#\bv} \lambda_2(\bv) \prod_{i\ne j} f(v_i,v_j) ( X(\bv) + \dots ) + \dots.
\end{multline}
Taking into account a consequence of the Bethe equations $\lambda_1(\bv) = \lambda_2(\bv)$ we instantly come to the equation \eqref{kat1}.
\qed

\subsection{Slavnov determinant formula}

 In this section we give a new way to get determinant representation of scalar product. 
It is based on theorem \ref{KAth}. 
Using this theorem one can writhe the following formula
\begin{equation}\label{sym-bu}
  S(\bx | \bu) =  \frac{1}{c^n}\mathop{{\rm Sym}}\limits_{\bar u} \; \prod_{i=1}^{n} g(u_i,x_i)
  \mathop{{\rm res}}\limits_{z=u_1} \tau(z|\bu)
  \mathop{{\rm res}}\limits_{z=u_2} \tau(z|\{x_1,\bu_1\}) \ldots
  \mathop{{\rm res}}\limits_{z=u_n} \tau(z|\{\bx_n,u_n\}),
\end{equation}
where $\#\bu=\#\bx=n$. 
Let us introduce
\be{jjk}
J(u_j,x_k)=\frac{c}{g(u_j,\bx)}\frac{\partial \tau(u_j|\bx)}{\partial x_k}=
\lambda_2(u_j)t(u_j,x_k)h(u_j,\bx)+(-1)^{n-1}\lambda_1(u_j)t(x_k,u_j)h(\bx,u_j).
\ee
Let $1\le m\le n$. Define
\be{nu}
\nu_k^{(m)}=\frac{\prod_{\substack{\ell=1\\ \ell\ne k}}^mg(x_k,x_\ell)}{\prod_{\ell=1}^m g(x_k,u_\ell)}.
\ee
Using a standard contour integral method one can easily prove that
\be{sumJm}
\sum_{k=1}^m J(u_j,x_k)\nu_k^{(m)}=-\frac{\mathop{\rm res}\limits_{z=u_j}\tau(z|u_1,\dots,u_m,x_{m+1},\dots,x_n)}
{c \prod_{\substack{i=1\\ i\ne j}}^m g(u_j,u_i)\prod_{\ell=m+1}^n g(u_j,x_\ell)},
\ee
for $1\le j\le m$. In particular, setting $m=1$ in \eqref{sumJm} we obtain
\be{sumJm1}
\tfrac1c\; g(u_1,x_1)\mathop{\rm res}\limits_{z=u_1}\tau(z|u_1,\bx_{1})=
J(u_1,x_1) g(u_1,\bx_1).
\ee
This formula, of course, can be also obtained via direct calculation.

 Let us make a replacement $u_j\to u_{n+1-j}$ and $x_j\to x_{n+1-j}$. Then equation \eqref{sym-bu} takes the form
\begin{equation}\label{sym-bu1}
  S(\bx | \bu) =  \frac{1}{c^n}\mathop{{\rm Sym}}\limits_{\bar u} \; \prod_{i=1}^{n} g(u_i,x_i)
  \mathop{{\rm res}}\limits_{z=u_i} \tau(z|\{u_1,\dots,u_i,x_{i+1},\dots,x_n\}).
\end{equation}

Now one can formulate theorem about determinant representation of scalar product.
\begin{theorem}\label{T-HNY}
Let $1\le m\le n$. 
Then scalar product between on-shell and off-shell Bethe vectors is
\begin{multline}\label{HNY}
  S(\bx | \bu) =  \frac{1}{c^{n-m}}\mathop{{\rm Sym}}\limits_{\bar u} \; \prod_{i=1}^{m}\left\{ J(u_i,x_i)\prod_{\ell=m+1}^{n}  g(u_i,x_\ell)\right\}
\Delta'(\{u_1,\dots,u_m\})\Delta(\{x_1,\dots,x_m\})   \\
 \times  \prod_{\ell=m+1}^{n}g(u_\ell,x_\ell)
  \mathop{{\rm res}}\limits_{z=u_\ell} \tau(z|\{u_1,\dots,u_\ell,x_{\ell+1},\dots,x_n\}).
\end{multline}
\end{theorem}

{\sl Proof.} We use induction over $m$. For $m=1$ the statement of the theorem is obvious. Indeed, using \eqref{sumJm1} we transform \eqref{sym-bu1} to the
form
\begin{equation}\label{HNY-m1}
  S(\bx | \bu) =  \frac{1}{c^{n-1}}\mathop{{\rm Sym}}\limits_{\bar u} \;  J(u_1,x_1)g(u_1,\bx_1)
  \prod_{\ell=2}^{n}g(u_\ell,x_\ell)
  \mathop{{\rm res}}\limits_{z=u_\ell} \tau(z|\{u_1,\dots,u_\ell,x_{\ell+1},\dots,x_n\}).
\end{equation}
This equation coincides with \eqref{HNY} at $m=1$.

Let \eqref{HNY} be valid for some $m\ge 1$. Then due to \eqref{sumJm} we have
\begin{multline}\label{sumJmp1}
\mathop{\rm res}\limits_{z=u_{m+1}}\tau(z|u_1,\dots,u_{m+1},x_{m+2},\dots,x_n)=
-c \prod_{i=1}^{m} g(u_{m+1},u_i)\prod_{\ell=m+2}^n g(u_{m+1},x_\ell)\\
\times\sum_{k=1}^{m+1} J(u_{m+1},x_k)\nu_k^{(m+1)},
\end{multline}
where $\nu_k^{(m+1)}$ are given by \eqref{nu} with $m$ replaced by $m+1$. Substituting this into \eqref{HNY} we find
\begin{multline}\label{HNY-mm}
  S(\bx | \bu) =  \frac{(-1)^{m+1}}{c^{n-m-1}}\mathop{{\rm Sym}}\limits_{\bar u} \; \Delta'(\{u_1,\dots,u_{m+1}\})\Delta(\{x_1,\dots,x_m\}) \prod_{i=1}^{m}J(u_i,x_i)
  \prod_{i=1}^{m+1}\prod_{\ell=m+1}^{n}  g(u_i,x_\ell)
  \\
 \times \sum_{k=1}^{m+1} J(u_{m+1},x_k)\nu_k^{(m+1)} \prod_{\ell=m+2}^{n}g(u_\ell,x_\ell)
  \mathop{{\rm res}}\limits_{z=u_\ell} \tau(z|\{u_1,\dots,u_\ell,x_{\ell+1},\dots,x_n\}).
\end{multline}
It is easy to see that the only term corresponding to $k=m+1$ survives in the sum over $k$ in the second line of \eqref{HNY-mm}. Indeed, for
all $k<m+1$ we obtain expressions of the form
\be{e-form}
J(u_{m+1},x_k)J(u_{k},x_k)\cdot G(u_k,u_{m+1}),\qquad\text{where}\qquad
G(u_k,u_{m+1})= - G(u_{m+1},u_k).
\ee
Such the contribution vanishes under symmetrization over $\bu$. Thus, we obtain
\begin{multline}\label{HNY-mm1}
  S(\bx | \bu) =  \frac{(-1)^{m+1}}{c^{n-m-1}}\mathop{{\rm Sym}}\limits_{\bar u} \Delta'(\{u_1,\dots,u_{m+1}\})\Delta(\{x_1,\dots,x_m\}) \prod_{i=1}^{m+1}J(u_i,x_i)
   \\
 \times \nu_{m+1}^{(m+1)}\prod_{i=1}^{m+1}\prod_{\ell=m+1}^{n}  g(u_i,x_\ell) \prod_{\ell=m+2}^{n}g(u_\ell,x_\ell)
  \mathop{{\rm res}}\limits_{z=u_\ell} \tau(z|\{u_1,\dots,u_\ell,x_{\ell+1},\dots,x_n\}),
\end{multline}
and substituting here
\be{nu1}
\nu_{m+1}^{(m+1)}=\frac{\prod_{\ell=1}^{m} g(x_{m+1},x_\ell)}{\prod_{\ell=1}^{m+1} g(x_{m+1},u_\ell)},
\ee
we immediately arrive at \eqref{HNY} with $m\to m+1$.\qed

\vspace{3mm}

Setting $m=n$ in \eqref{HNY} we obtain
\begin{equation}\label{HNY-n}
  S(\bx | \bu) =  \mathop{{\rm Sym}}\limits_{\bar u} \;\Delta'(\bu)\Delta(\bx) \prod_{i=1}^{n}J(u_i,x_i),
\end{equation}
and thus,
\begin{equation}\label{HNY-res}
  S(\bx | \bu) =  \Delta'(\bu)\Delta(\bx)\det_n J(u_i,x_k).
\end{equation}
We can rewrite equation \eqref{HNY-res} in the following form
\begin{equation}
   S(\bx | \bu) =  \Delta'(\bu)\Delta(\bx)\det_n \frac{c}{g(u_j,\bx)}\frac{\partial \tau(u_j|\bx)}{\partial x_k}.
\end{equation}
and it is also known as Slavnov determinant formula \cite{Sla89} for scalar product between on-shell and off-shell Bethe vectors.

\section{Reflection case}\label{reflect}

\subsection{Bethe vector}

 Now we pass to the description of Bethe vectors in onen boundary case.
They belong to the same Hilbert space  $\mathcal{H}$.
Basic of the space is generated by multiple action of $\mT_{12}$ onto vacuum $\rvac$.
According to the \cite{Skl88} in the framework of generalized model off-shell Bethe vector for set $\bu = \{u_1, \ldots, u_n\}$ is  
\begin{equation}\label{rbv}
 \hBB(\bu) = \mT_{12}(\bu) \rvac = \mT_{12}(u_1)\ldots \mT_{12}(u_n) \rvac.
\end{equation}
It is easy to prove that the Bethe vector \eqref{rbv} is symmetric over permutations of the parameters $\bu$ \cite{Skl88}.

 One can prove that off-shell Bethe vector becomes eigenvector of transfermatrix $\hat t (u)$ \eqref{rtm} 
\begin{equation}
 \hat t(z) \, \hBB(\bu) = \hat\tau(z | \bu) \, \hBB(\bu)
\end{equation}
if parameters $\bu$ satisfy the system of Bethe equations \cite{Skl88}
\begin{equation}\label{rbe}
 \frac{\left(u_i + \xi_+ - \frac{c}{2} \right)\left(u_i + \xi_- - \frac{c}{2} \right) }{\left(u_i - \xi_+ + \frac{c}{2} \right)\left(u_i - \xi_- + \frac{c}{2} \right)}
 \frac{\lambda_1(u_i)\lambda_2(-u_i)}{\lambda_1(-u_i)\lambda_2(u_i)} = \frac{f(u_i,\bu_i)f(u_i, -\bu_i)}{f(\bu_i,u_i)f(-\bu_i,u_i)},
\end{equation}
with eigenvalue 
\begin{equation}\label{rtme}
\begin{aligned}
 \hat\tau(z | \bu) = & \frac{2z + c}{2z}\left( z + \xi_+ - \frac{c}{2} \right)\left( z + \xi_- - \frac{c}{2} \right) f(\bu,z)f(-\bu,z) \lambda_1 (z)\lambda_2 (-z) + \\ 
 & \frac{2z - c}{2z}\left( z - \xi_+ + \frac{c}{2} \right)\left( z - \xi_- + \frac{c}{2} \right) f(z,\bu)f(z,-\bu) \lambda_1 (-z)\lambda_2 (z).
\end{aligned}
\end{equation} 
 
 Dual Bethe vectors belong to the dual space $\mathcal{H}^*$.
They can be obtained as a transposition of the Bethe vectors \cite{Skl88}.
We denote dual Bethe vectors by $\hCC(\bu)$. It can be constructed in the similar way \cite{Skl88} using dual vacuum $ \lvac$ from the dual space $\mathcal{H}^*$ with the properties 
\begin{equation}
 \begin{aligned}
 \lvac \mathcal{T} _{11}(u) &=   \left( u + \xi_-  - \frac{c}{2} \right) \lambda_1(u)\lambda_2(-u)  \lvac, \\
 \lvac \mathcal{T} _{22}(u) &= \left(\frac{c-2u}{2u}\left( u- \xi_-  + \frac{c}{2} \right) \lambda_1(-u)\lambda_2(u) + \frac{ u+ \xi_- - \frac{c}{2}}{2u}  \lambda_1(u)\lambda_2(-u) \right)  \lvac, \\
 \lvac \mathcal{T} _{12}(u)&= 0.
 \end{aligned}
\end{equation} 
Dual Bethe vector becomes on-shell too, if the set $\bu$ satisfy the system of Bethe equations \eqref{rbe} with the same eigenvalue \eqref{rtme}.

\subsection{Reflection main theorem and Slavnov determinant formula}

 In this subsection we present the main results of the paper for open boundary boundary case (case of reflection algebra). 
Here we give two theorem \ref{rKA} and \ref{rSl}. 
Proofs of the theorems are the same as in periodic case.
This theorem  allows us to calculate scalar product of on-shell and off-shell Bethe vectors for open boundary boundary case, knowing the action of the reflection transfer matrix \eqref{rtm} onto a Bethe vector \eqref{rbv}.

 Let us recall the formula of action of reflection transfer matrix \eqref{rtm} onto Bethe vector \eqref{rbv} \cite{Skl88}
\begin{equation}
   \hat t(z) \; \hBB(\bu) =   
  \hat\tau(z|\bu) \,\hBB(\bu) -  \frac{1}{c}  
  \sum_j \frac{2 u_i}{c + 2 u_i} \frac{c + 2 z}{c} g(z, u_j) g(z,- u_j) \mathop{{\rm res}}\limits_{w=u_j} \hat\tau(w|\bu)
 \; \hat\BB(\{z, \bu_j \}).
\end{equation}
We see that the action is given only by eigenvalue of transfer matrix  $\hat\tau$ \eqref{rtme}.

\begin{theorem}\label{rKA} 
Let us consider multiple action of a transfer matrix onto off-shell Bethe vector with  $\#\bu = \#\bv$
 \begin{equation} 
   \hat t (\bv) \; \hBB(\bu) = \sum \hat R(\bv\,|\,\bet_{\so},\bet_{\st}) \; \hBB(\bet_{\st}),
 \end{equation}
where $\bar\eta=\{\bu,\bv\}$. 
Here the sum is taken over partitions $\eta$ into two subsets $\eta_{\so}$ and $\eta_{\st}$, such that  $\#\eta_{\so} = \#\eta_{\st}$.

Then the coefficient of $\hBB(\bv)$ of multiple action of the transfer matrixes  with $\#\bv=\#\bu$ coincides with the scalar product of on-shell--off-shell Bethe vectors
 \begin{equation}\label{kat1}
  S(\bv | \bu) =\hat R(\bv\,|\,\bu ,\bv).
 \end{equation}
 where set $\bv$ satisfies Bethe equations and set $\bu$ is free (off-shell).
\end{theorem}
Proof of the theorem is absolutely the same as for periodic case.

 Using way of derivation similar to periodic case one can get determinant representation of scalar product between eigenvector of reflection transfer matrix \eqref{rtm} and off-shell Bethe vector \eqref{rbe}
\begin{theorem}\label{rSl}
Let us consider multiple action of a transfer matrix onto off-shell Bethe vector with  $\#\bu = \#\bv$
\begin{equation}\label{rSl}
   S(\bx | \bu) =  \Delta'(\bu^2)\Delta(\bx^2)\det_n \frac{2 x_k - c}{2 x_k}\frac{c}{ (2 u_j -c)} \frac{c }{g(u_j,\bx)g(u_j,-\bx)}\frac{\partial \tau(u_j|\bx)}{\partial x_k}.
\end{equation}

\end{theorem}

One can compare formula \eqref{rSl} with \cite{Kit07}.

\section*{Conclusion}

 In this paper we formulate and prove relation between action of the transfer matrix onto Bethe vectors and scalar product of Bethe vectors. 
The main results of the paper are theorems \ref{KAth} and \ref{rKA}.
This way certainly is more direct and simple than the methods used before for the derivation of the scalar product of on-shell and off-shell Bethe vectors.
This result completely matches to the well-known Slavnov determinant formulas in periodic and reflection boundary cases.
The last one was not known in general case of reflection algebraic Bethe ansatz but only in spin chain case.  

 In both cases we have determinants of matrices constructed by derivatives of the eigenvalues of transfer matrices.
This circumstance was known for a long time, but without explanation.
We give this explanation. 
Action of the transfer matrix onto Bethe vector is given only by the function of eigenvalue of transfer matrix \eqref{tme}. 
And taking to account our results it is natural to expect that the scalar product is given only by the eigenvalue \eqref{tme} too.

 There is a direct generalization of the our relation to higher rank algebraic Bethe ansatz (also known as nested Bethe ansats). 
For example, in the case of $\mathfrak{gl}_3$ algebra in terms of paper \cite{Bel12} action the transfer matrix is closely related with scalar product too. More precisely, in the multiple action of the transfer matrix onto Bethe equation 
\begin{equation}
 t(\buc) t(\bvc) \BB(\bub,\bvb)
\end{equation}
coefficient of the  $\BB(\buc,\bvc)$ can be identified with the scalar product of on-shell and off-shell Bethe vectors as in the current paper.
A proof of this statement is the similar to the proof for $\mathfrak{gl}_2$ given in this work.
But at the moment there is no determinant representation for such scalar product. 
We hope that out approach will help to get such determinant representation even for higher rank case in the future.

\section*{Acknowledgments}
We are grateful to Arthur Hutsalyuk for inspiring discussions and to Nikita Slavnov for fruitful discussions. 
The work of A.L. has been funded.

\end{document}